\documentstyle[aps,prl,twocolumn,epsf,graphics,psfig]{revtex}

\begin{document}
\twocolumn[\hsize\textwidth\columnwidth\hsize\csname @twocolumnfalse\endcsname
\title{Infrared conductivity of metallic (III,Mn)V ferromagnets} 
\author{Jairo Sinova$^{1}$, T. Jungwirth$^{1,2}$, S.- R. Eric Yang $^{1,3}$,
J. Ku\v{c}era$^{2}$, and A.H. MacDonald$^{1}$}
\address{$^{1}$Department of Physics,
The University of Texas at Austin, Austin, TX 78712 \\}
\address{$^{2}$Institute of Physics ASCR, Cukrovarnick\'a 10,
162 53 Praha 6, Czech Republic\\}
\address{$^3$Department of Physics, Korea University, Seoul 136-701, Korea}
\date{\today}
\maketitle
\begin{abstract}
We present a theory of the infrared conductivity and 
absorption coefficients of metallic (III,Mn)V ferromagnetic semiconductors.
We find that the conductivity is dominated by inter-valence-band 
transitions that produce peaks at $\hbar \omega \sim 220 {\rm meV}$ and obscure
the broadened Drude peak.  We demonstrate that transverse f-sum rule measurements 
can be used to extract accurate values for the free carrier density, 
bypassing the severe characterization difficulties that 
have till now been created by the large anomalous Hall effect in these materials.
\end{abstract}


\vskip1pc]

Following the discovery of carrier-induced ferromagnetism in (In,Mn)As
\cite{ohnoprl92} and (Ga,Mn)As \cite{ohnoapl96} in the early nineties, 
studies of ferromagnetism in (III,Mn)V diluted magnetic semiconductors (DMS's)
have yielded many surprises.  The transport and optical properties 
of these heavily-doped semiconductors are richer
than those of conventional itinerant electron ferromagnets
because of strong valence-band spin-orbit coupling, and because
of the sensitivity of their magnetic state to 
growth conditions, doping, and external fields.  These novel ferromagnets are 
likely to have a major technological impact \cite{dietlacta01} if systems with 
Curie temperatures above room temperature can be created and better 
control of disorder effects can be achieved.  
We present a theory 
of the infrared conductivity of (III,Mn)V ferromagnets which we believe will  
increase the value of this powerful probe in characterizing present and future DMS ferromagnets. 
In particular, we demonstrate that infrared conductivity measurements in metallic samples
can be used to obtain accurate measurements of the total free carrier density, likely 
the key to understanding the now firmly established\cite{schiffer} dependence
of magnetic properties on growth and post-growth annealing conditions.
This experimental possibility is especially important in these semiconductors because they
are ferromagnetic and have a large anomalous Hall conductivity \cite{jungwirthahe} which
severely complicates Hall-effect carrier-density measurements, \cite{ohnoapl96,ohnojmmm99} 
even when they are performed in very strong magnetic fields. 

Our conclusions are based on a semi-phenomenological model
\cite{dietlsci00,dietlprb01,abolfathprb01,bookchapter}
in which host semiconductor valence band 
electrons, described by a ${\bf k.p}$ model, interact with 
randomly distributed aligned ($T=0$) Mn local- moment acceptors via Coulomb 
and short-range exchange interactions,
and via Coulomb interactions only with the compensating charged defects known to be present.
Some of our considerations are based on standard linear-response theory expressions 
for the ac conductivity of weakly disordered metals, in which disorder is included
only through finite quasiparticle lifetimes and localization effects are ignored.
In the dc limit, with quasiparticle scattering rates evaluated using
Born approximation expressions for scattering off the spatial fluctuations 
of the screened-Coulomb and exchange potentials, these expressions imply mobilities \cite{ourdcpaper}
that are consistent with values measured in optimally
annealed, high-$T_c$, DMS ferromagnets.  Although the predictions discussed below
are intended to be most reliable for the most metallic systems, they appear to explain
much currently available infrared optical data, most of which 
has been obtained in studies of systems with relatively low dc conductivities. 

The main features observed in experiments \cite{nagai,Katsumoto,HirakawaGaAs,SanDiego} are 
the absence of a clear low-frequency Drude peak, 
and a broad absorption peak near $220$ meV that becomes stronger as the samples are cooled.
The $220$ meV peak has been attributed either to inter-valence-band transitions,
to transitions between the semiconductor valence band states and a Mn induced impurity band,
or to a combination of these contributions.  Dynamical mean-field-theory studies \cite{Hwang}, 
for a single-band model that neglects the spin-orbit coupling and the heavy-light degeneracy 
of a III-V semiconductor valence band, demonstrate that non-Drude impurity-band related features 
can occur in DMS ferromagnets. The conductivities 
predicted by this model are, however, inconsistent with experiment in magnitude 
and in temperature trend.  As demonstrated in the following paragraphs, realistic features of the 
valence band electronic structure are key to understanding the conductivity data.
We find that the absorption peak near 220~meV is associated
with heavy-hole to light-hole inter-valence-band transitions,
and that transitions to the bands split-off by spin-orbit interactions  
are responsible for the gradual increase in absorption observed in 
experiments \cite{Katsumoto} at frequencies above $500$ meV. 

\begin{figure}
\epsfxsize=3.20in
\centerline{\epsffile{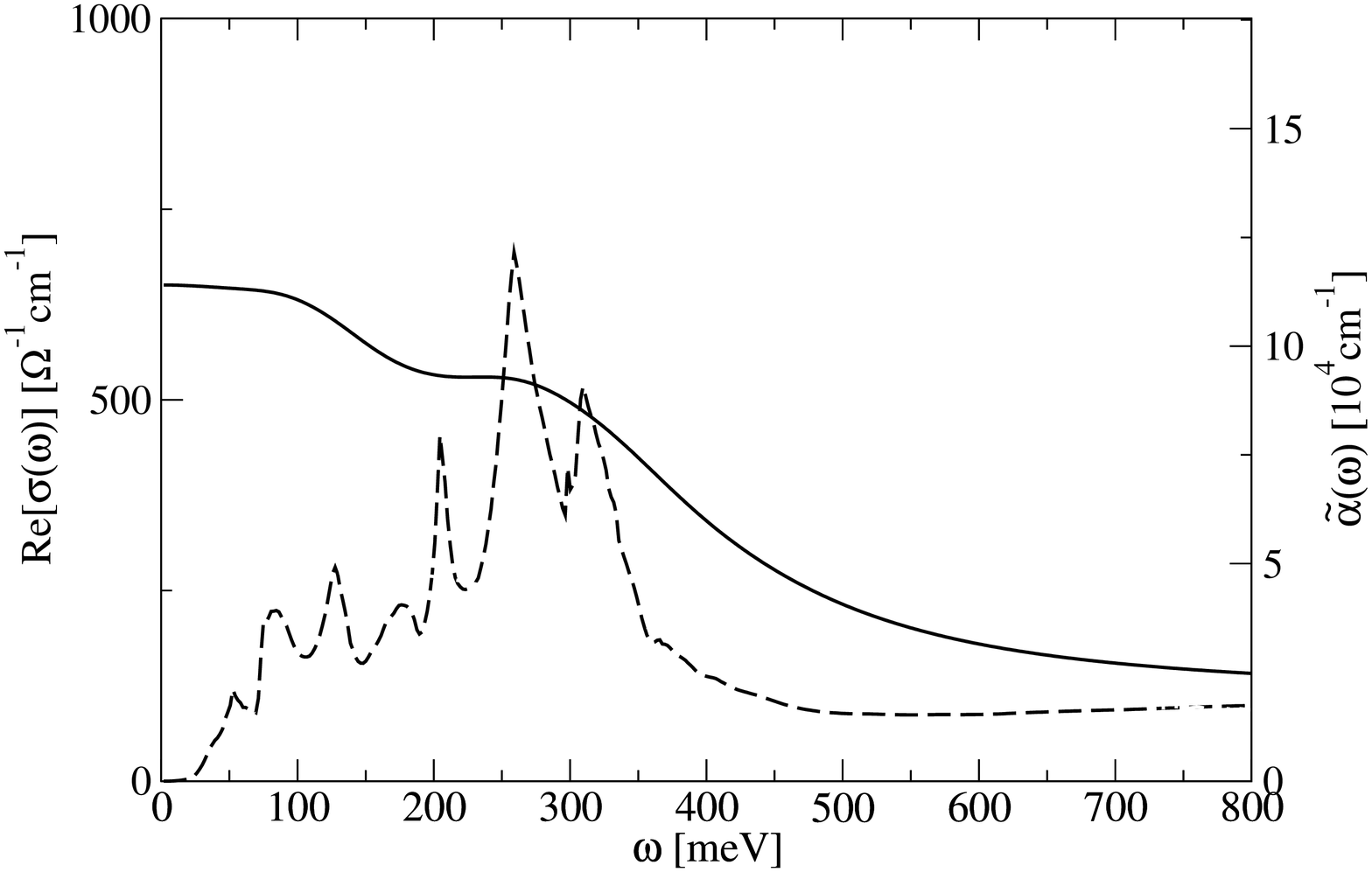}}
\caption{Optical conductivity Re[$\sigma(\omega)$] and absorption 
coefficient $\tilde{\alpha}(\omega)$
for carrier density $n=0.3 {\rm nm}^{-3}$
in Ga$_{0.95}$Mn$_{0.05}$As without (dashed) and with (solid)  
impurity scattering present.}
\label{fig1}
\end{figure}

The physical picture we have developed for the infrared conductivity starts 
from an examination of the limit in which disorder is ignored.  In building a picture
starting from this limit we are motivated in part by
the observation that many magnetic and transport property
trends (critical  temperatures, magnetic anisotropy and 
its strain dependence, magnetic stiffness, anomalous Hall effect, etc.
\cite{dietlsci00,dietlprb01,abolfathprb01,bookchapter,dietlprb97,jungwirthprb99})
in DMS ferromagnets appear to be adequately described without accounting
for disorder, even though it is relatively strong by some measures. 
The linear response of the system to a
light probe is characterized by the optical conductivity tensor $\sigma_{\alpha,\beta}(\omega)$.
In thin film absorption measurements, the real part of the conductivity is 
related to the absorption coefficient by
\begin{equation}
\tilde{\alpha}(\omega)=2\frac{{\rm Re[}\sigma(\omega)]}{Y+Y_0}\,\,,
\end{equation}
where $Y$ and $Y_0$ are the admittances of the substrate and free space, respectively \cite{note_alpha}.
We have evaluated $\sigma_{\alpha,\beta}(\omega)$ at $T=0$ using the standard Kubo formula
for non-interacting quasi-particles\cite{Kubo}.  The clean-limit ac conductivity of our model
is plotted in Fig.~\ref{fig1} for a typical set of (Ga,Mn)As parameter values, carrier density 
$p=0.3$~nm$^{-3}$, 5\% Mn substitution, implying a Mn concentration $N_{Mn}=1.1$~nm$^{-3}$,
and band parameters that incorporate the strain 
produced by epitaxial growth on a GaAs substrate \cite{msweb}.  When disorder is neglected,
Coulomb interactions have no effect, while the exchange interactions produce a uniform 
exchange mean Zeeman field, $\vec h = J_{pd} N_{Mn} S \hat \Omega$, where $\hat\Omega$
is the magnetization direction.
The finite-frequency conductivity results from 
transitions between occupied and unoccupied valence bands states; the band 
Kramers degeneracies are split by an amount 
comparable to the Fermi energy.  ($J_{pd}=55$~meV~nm$^3$ is the exchange-coupling constant 
so that $h \sim 150 {\rm meV}$.)  The results illustrated in Fig.~\ref{fig1} are for 
$\hat \Omega$ along the  growth ($\langle 001\rangle$) direction, the magnetic 
easy-axis for a tensile strained (Ga,Mn)As layers \cite{abolfathprb01}; 
clean limit results for samples under compressive strain (GaAs substrates) 
which have in plane easy axes differ in detail \cite{msweb}.
The two split-peaks present in the clean limit at $\sim 250$ and $\sim 325$ meV correspond to
heavy-hole to light-hole transitions and to light-hole to split
off (s-o) band transitions, respectively. Beyond $\sim 500$ meV there is a slight rise 
in the absorption which corresponds to the heavy-hole to  
s-o split off band transition.  In the clean limit, the conductivity has a 
$\delta$-function contribution at $\omega = 0$ due to intraband transitions.
Its relative contribution to the f-sum rule of our model, discussed further below, 
is 35\%, nearly independent of carrier density over the relevant range.

\begin{figure}
\epsfxsize=3.20in
\centerline{\epsffile{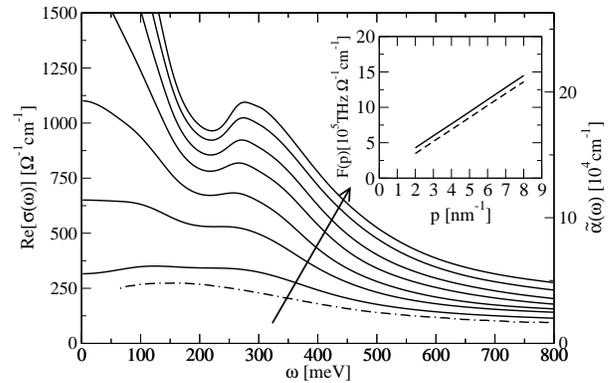}}
\caption{Optical conductivity ${\rm Re}[\sigma(\omega)]$ and absorption 
coefficient $\tilde{\alpha}(\omega)$
for carrier densities from $p=0.2$ to $0.8 {\rm nm}^{-3}$ in
the direction indicated by the arrow, 
for Ga$_{0.95}$Mn$_{0.05}$As. The dot-dashed line is an experimental
absorption curve for a sample with a Mn concentration of 4\% obtained by 
Hirakawa {\it et al}. The inset shows the spectral weight $F(p)$ 
evaluated with a frequency cut-off of 800 meV including disorder (dashed) and in
the clean limit (solid).}
\label{fig3}
\end{figure}

We account for disorder by including the lifetime broadening of quasiparticle
spectral functions in evaluating the Kubo formula.   
The effective lifetime for transitions between bands $n$ and $n^{\prime}$, 
$\tau_{n,n^{\prime}}$, is calculated  
by averaging quasiparticle scattering rates calculated from Fermi's golden rule including both 
screened Coulomb and exchange interactions \cite{ourdcpaper}.
The solid curve in Fig.~\ref{fig1} illustrates the influence of disorder on the 
conductivity.  The intraband
conductivity evolves into a broadened Drude peak that overlaps with broadened 
interband absorption features.  In Fig.~\ref{fig3}, we plot ac conductivities 
calculated in this way for ${\rm Ga}_{.95}{\rm Mn}_{.05}$As at a series of carrier
densities.  In our model, an increase in carrier density is accompanied by 
a decrease in the density $N_c$ of compensating charged defects;  
we assume that all have charge $Z=2$ so that $p=N_{Mn}- 2N_{c}$ \cite{note3}.
As $p$ increases, $N_{c}$ and quasiparticle
scattering rates decrease in tandem.  The evaluation of quasiparticle scattering rates  
allows us to check the consistency of our weak-scattering approach, which  requires 
$\hbar/(\tau_{nn'} E_F)>1$, where $E_F$ is the hole Fermi energy.
For 5\% Mn doping, the condition is safely satisfied only for $p \gtrsim 0.3
{\rm nm}^{-3}$.  As the carrier density increases above this value we 
see in Fig.~\ref{fig3} that the peak near 220 meV sharpens and the low frequency Drude peak
emerges more clearly.

Since all model parameters are accurately known
from other experiments, the ac conductivity values in Fig.~\ref{fig3} are
parameter free.  
Although direct comparisons with infrared conductivity experiments 
is difficult because the carrier density is not usually known, our
dc conductivities are generally larger than those of optimally annealed samples 
\cite{schiffer}. 
Our quasiparticle scattering rates are likely smaller
than those in any current experimental sample, due to some combination of 
inaccuracy in our approximate scattering amplitudes and unknown 
sources of additional unintended disorder.  Further progress in sorting
out variable transport and magnetic properties requires accurate
experimentally determined carrier density 
values, which we believe can be obtained directly from infrared conductivity data 
as we explain below.  

In the following paragraphs we discuss the f-sum rule of our model:
\begin{equation}
F \equiv \int_0^{\infty} d\omega {\rm Re}[\sigma_{xx}(\omega)]=
\frac{\pi e^2}{ 2V} \sum_{\alpha} f_{\alpha} \;\;
\langle \alpha | \frac{\partial^2 H_{band}} {\hbar^2 \partial k_x^2} | 
\alpha\rangle.
\label{eq:realfsumrule} 
\end{equation}
In this equation $f_{\alpha}$ is a quasiparticle Fermi factor and 
$\partial^2 H_{band}/ \hbar^2 \partial k_x^2$ is the $xx$ component of the 
${\bf k.p}$ model inverse effective mass operator, which is diagonal in 
envelope function wavevector but off diagonal in envelope function spinor index.
We emphasize that this sum rule is completely independent of the weak-scattering
approximations on which the ac conductivity calculations discussed above were 
based.  In practical applications, it is necessary to choose an 
upper cut-off for the frequency integral \cite{note2}.
We define a weakly cut-off dependent optical effective mass for DMS ferromagnets
by defining $F=\pi e^2 p / 2 m_{opt}$.
As illustrated in the inset of Fig \ref{fig3}, for the case of cut-off frequency  
$\hbar\omega_{max}=800 {\rm meV}$,
the f-sum rule values of $F$ evaluated from our weak scattering theory are 
accurately linear in $p$ over the entire range of relevant carrier densities.
The optical masses evaluated here, which provide the f-sum rule with practical
utility, represent a complicated average over heavy and light hole masses in the 
spin-split ferromagnetic valence bands.
Disorder does have a small but measurable effect on $m_{opt}$ as illustrated 
in Fig.~\ref{fig3}.  By solving self-consistent mean-field calculations 
for finite-size realizations of our DMS ferromagnet model, we have established that 
a fully non-perturbative treatment of disorder alters $m_{opt}$, by less than 10\% for 
metallic DMS ferromagnets\cite{ericpreprint,Eric}. 
We have also verified that the optical mass is not perceptibly dependent on the 
magnetization orientation and that it is altered by less than 1\% by the exchange mean-field.
Using the six-band Luttinger model, the optical masses for
GaAs, InAs, and GaSb DMS ferromagnets with a 800, 400, and 700 {\rm meV} cutoffs
are 0.25-0.29 m$_e$, 0.40-0.43 m$_e$, and 0.21-0.23 m$_e$ respectively, the extremes
of these ranges corresponding to the clean (lower) and disordered (upper) limits 
of our model \cite{note4}.  
{\em $m_{opt}$ depends to a good approximation only on the frequency cutoff and 
on the Luttinger parameters that characterize the host semiconductor valence bands \cite{iiiv}, 
permitting accurate carrier-density measurements based on the f-sum rule.} 

We have applied our procedure for determining the DMS carrier density to
a sample with 4\% Mn doping, measured by 
Katsumoto {\it et al} \cite{Katsumoto}, 
obtaining a value of 0.17 $\pm .03$ nm$^{-3}$
(Their data is shown as the dot-dashed line in Fig. \ref{fig3}).
Although their sample is strongly disordered, the 
f-sum rule based density estimate still appears reasonable. It is consistent, 
within 30\%,  with the 
\cite{msweb,jungwirth0201157} carrier density that
would produce a ferromagnetic state with the measured \cite{Katsumoto}
critical temperature of 
90~K. Note that the hole density estimate \cite{Katsumoto} based on the measured
room temperature Hall coefficient for the same sample is $\sim 5\times$ smaller than 
that expected on the basis of its critical temperature, emphasizing once again
the difficulty of extracting carrier densities from Hall data.  

\begin{figure}
\epsfxsize=3.00in
\centerline{\epsffile{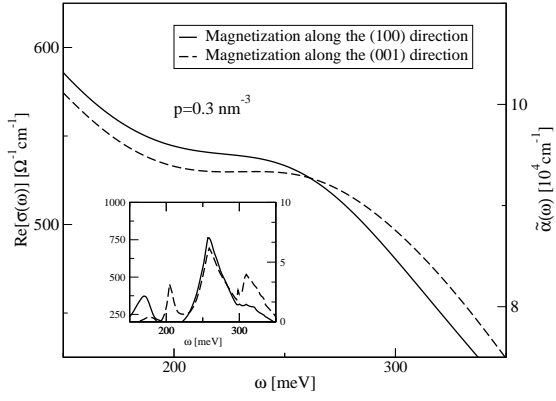}}
\caption{Magnetization orientation dependence of the optical
conductivity ${\rm Re}[\sigma(\omega)]$ and absorption 
coefficient $\tilde{\alpha}(\omega)$
for carrier densities $n=0.3 {\rm nm}^{-3}$
for Ga$_{0.95}$Mn$_{0.05}$As at $T=0$.
The light is linearly polarized along the x-direction.
The inset indicates the results for the disorder free case.}
\label{fig2}
\end{figure}

The strong spin-orbit coupling in the (Ga,Mn)As valence band
($\Delta_{so}=341$~meV) compared to the typical spectral broadening of the
hole quasiparticles ($\hbar/\tau\sim 100-200$~meV)
leads to a dependence of the
absorption spectrum  on the direction of the magnetization.
This dependence should lead to measurable infrared conductivity changes that
can be induced by an external
magnetic field larger than the sample's magnetocrystalline
anisotropy field $\sim 100$~mT \cite{dietlprb01},  applied along the
hard $\langle 100\rangle$ magnetic  axis.
The inset of Fig.~\ref{fig2} shows the orientation dependence of the optical conductivity
and absorption coefficients for the same sample parameters as
in Fig.~\ref{fig1}, calculated neglecting disorder and with 
the incident light polarization along the $\langle 100\rangle$ direction.
In the main plot of Fig.~\ref{fig2} we 
compare absorption spectra evaluated using our weak disorder approximation.
The sharp spectral features
near 325~meV are smeared out by disorder, however, the $\sim$15~meV
relative shift of the absorption peak for the two magnetization
orientations is nearly disorder independent and should be
observable experimentally.
This effect is stronger when the density of compensating defects is reduced.

In summary, we find that the infrared conductivity of metallic DMS ferromagnets has 
Drude and interband contributions that have approximately equal importance.
In current samples these contributions overlap in frequency, but we expect that they
will begin to separate if recent improvements in sample conductivity can be extended. 
We predict that the infrared conductivity will be very sensitive to changes in 
compensating defect and carrier densities, believed to be responsible for the sensitivity
of DMS ferromagnets to growth and annealing conditions.  In particular, we have evaluated
optical effective masses that will allow carrier densities to be extracted from
f-sum rule measurements, circumventing difficulties posed by the large anomalous Hall effect 
of (III,Mn)V ferromagnetic semiconductors.  Simultaneous measurement of
optical, magnetic, and dc transport properties should prove to be potent 
in advancing the materials science of these interesting systems, when our optical 
masses are used to extract carrier densities.
We have also demonstrated the sensitivity of the infrared optical 
conductivity to magnetization orientation that is due to the strong
spin-orbit coupling in these systems. 

We acknowledge helpful discussions with Jacek Furdyna, Bruce McCombe, 
Andrea Markelz, Joaquin Fernandez-Rossier, Peter Schiffer, and Jason Singley.
The work was supported by
the Welch Foundation, DARPA, the Grant Agency of the Czech Republic
under grant 202/02/0912, the Ministry of Education of the Czech Republic
under grant OC P5.10, and KOSEF Quantum-functional
Semiconductor Research Center at Dongguk University.

\begin{references}
\bibitem{note_alpha}  This formula is applicable in the far infrared regime. At higher 
frequencies in the spectrum multiple scattering from the interfaces have to be
taken into account. This can be easily done using the standard electromagnetic formalism
of light waves propagating through a medium.
Note that $\tilde{\alpha}(\omega)$ is not the same as the bulk absorption coefficient 
usually defined as $\alpha(\omega)={2\omega{\rm Im} [\sqrt{\epsilon(\omega)}]}/{c}$.
\bibitem{note2}
Note that Eq.~\ref{eq:realfsumrule}, is useful only when the maximum frequency
of the intra-valence-band infrared conductivity is well below
the sum of the semiconductor gap and the Fermi energy. 
This conditions is reasonably well satisfied for most DMS ferromagnets.
Note also that the application of a frequency cutoff to such equation necessarily increases
the $m_opt$ measured.
\end{references}
\end{document}